\documentclass[a4paper,12pt]{article}

\usepackage[margin=1in]{geometry}
\usepackage{hyperref}
\usepackage{amsmath}
\usepackage{amssymb}
\usepackage{multirow}
\usepackage{booktabs}
\usepackage{graphicx}

\usepackage{natbib}

  \title{\textbf{Bayesian beta nonlinear models with constrained parameters to describe ruminal degradation kinetics\footnote{The original article is published 
  			in Journal of Applied Statistics. Published online: 09 Apr 2021, https://doi.org/10.1080/02664763.2021.1913105. This reprint differs from the original in pagination and typographic detail.}}}

\begin{document}
\author{
D. Salmer\'on\\ 
\small{Departamento de Ciencias Sociosanitarias, IMIB-Arrixaca,} \\
\small{Universidad de Murcia, Murcia, Spain}.\\
\small{CIBER de Epidemiolog\'ia y Salud P\'ublica (CIBERESP), Madrid, Spain.}
}

\maketitle

\begin{abstract}
	The models used to describe the kinetics of ruminal degradation are usually nonlinear models where the dependent variable is the proportion of degraded food. The method of least squares is the standard approach used to estimate the unknown parameters but this method can lead to unacceptable predictions. To solve this issue, a beta nonlinear model and the Bayesian perspective is proposed in this article. The application of standard methodologies to obtain prior distributions, such as the Jeffreys prior or the \textit{reference priors}, involves serious difficulties here because this model is a nonlinear non-normal regression model, and the constrained parameters appear in the log-likelihood function through the Gamma function. This paper proposes an objective method to obtain the prior distribution, which can be applied to other models with similar complexity, can be easily implemented in OpenBUGS, and solves the problem of unacceptable predictions. The model is generalized to a larger class of models. The methodology was applied to real data with three models that were compared using the Deviance Information Criterion and the root mean square prediction error. A simulation study was performed to evaluate the coverage of the credible intervals.
\end{abstract}


\section{Introduction} 
Feed ingested by ruminant animals is subjected to degradation in the rumen, and the final products of the degradation are used for the synthesis of microbial biomass. The nutritional value of a feed depends on its nutrient content, the extent of degradation and the digestibility of non-degraded food components. To evaluate the nutritional status of ruminant animals and to predict the amount of nutrients required, the accurate estimation of the degradation of the feed that they receive is paramount. 

The model proposed by \O rskov and McDonald \cite{OK1979} is widely used to describe the kinetics of ruminal degradation and to provide information on the quality and nutritional characteristics of food. According to the main collection of the Web of Science, this model has been cited by at least 3029 research articles. The model assumes that $y(t)$, the proportion of degraded food up to time $t$, is given by the nonlinearizable curve $y(t)=a+b(1-e^{-ct})$, where $a$ means the proportion of degraded food almost instantaneously and $b$ means the remaining proportion of food to be degraded at a velocity that is controlled by $c$. Due to the biological meaning of these parameters, they
have to satisfy the following constraints: $a\in(0,1)$, $b\in(0,1)$, $a+b\in(0,1)$, and $c>0$.

The method of least squares is the standard approach used to estimate the unknown parameters $a$, $b$, and $c$. This method presents two deficiencies for this problem. First, the distributions of the estimators are unknown, and hence, asymptotic approximations are usually used in standard statistical packages to obtain standard errors and confidence intervals. However, very often, the sample size is not large enough in the experiments performed to study degradation kinetic curves, which limits the application of asymptotic approximations. Second, and perhaps more important, the method can lead to unacceptable predictions if the estimates do not satisfy the above constraints on $a$, $b$, and $c$. For instance, the data $\boldsymbol{y}=(0.38,0.51,0.59,0.79,0.89)$, $\boldsymbol{t}=(3,6,9,15,24)$, have been obtained from \cite{OK1979}, the unrestricted maximum likelihood 
estimates of $a$, $b$, and $c$ are 
$\hat{a}=0.21001$, $\hat{b}=0.8270$, and $\hat{c}=0.0742$, respectively, and the frequentist 95\% confidence intervals are  
(-0.0178, 0.4380), (0.5336, 1.1210), and (-0.0164, 0.1648), respectively. The confidence intervals contain inadmissible values, $\hat{a}+\hat{b}=1.03701$, and for $t>41.7$, the prediction of the proportion of food degraded is $\hat{y}=\hat{a}+\hat{b}(1-\exp(-\hat{c}t))>1$.

Cano and Salmer\'on \cite{CanoSalmeron2007} have shown that the Bayesian approach avoids
these disadvantages automatically if the constraints are taken into account in the prior
distribution and consequently in the
posterior distribution. On the other hand, the Bayesian model in \cite{CanoSalmeron2007} assumes that the distribution of the observed proportions is the normal distribution; concretely, if $y_i$ is the observed proportion at times $t_i$, $i=1,\dots,n$, then the model assumes that 
\begin{equation}
	y_i\mid a,b,c,\sigma\sim N(a+b(1-e^{-ct_i}),\sigma^2),\,\,\,i=1,\dots,n. \label{modelo_normal}
\end{equation}
However, the data are proportions, and hence, the normal distribution might not be suitable. A reasonable solution to improve the model could be to apply a transformation, such as the logit or the log-log, and then to assume that the distribution of the transformed data is normal. However, this procedure does not allow imposing that the mean of $y_i$ is $a+b(1-e^{-ct_i})$ in a treatable way. Suppose that the transformation is $y_i=H(z_i)$ and that the model for $z_i$ is $z_i\sim N(\theta_i,\sigma_z^2)$. To obtain
\[
a+b(1-e^{-ct_i})=\int_{-\infty}^{+\infty}H(z)N(z\mid\theta_i,\sigma_z^2)\mathrm{d}z,
\]
we would need to solve $(\theta_i,\sigma_z^2)$ as a function of $(a,b,c)$, which complicates the inference procedure. The solution adopted in this article models the proportion $y_i$ using the beta distribution.

The  proposed model is a beta nonlinear regression model, and it is presented in section 2. The model takes advantage of the parameterization used in \cite{Ferrari2004}, but the mean of the response variable, $y_i$, is expressed as $a+b(1-e^{-ct_i})$ to preserve the interpretation of the parameters instead of using the logit link function as in \cite{Ferrari2004}. In section 3, a prior distribution is proposed that takes into account the interpretation of the parameters, and how the Bayesian model can be implemented in OpenBugs is demonstrated. Although the model of \O rskov and McDonald is widely used, these results are generalized to a large class of models in section 4. Section 5 is dedicated to illustrating the procedure with real and simulated data.

\section{The beta nonlinear regression model}

Let $y_1,\dots,y_n$ be the independent observed proportions of degraded food at
times $t_1,\dots,t_n$, where $0<t_1\leq \dots\leq t_n$. The beta distribution $\mathcal{B}(p,q)$ with density function
\[
f(y\mid p,q)=\frac{\Gamma(p+q)}{\Gamma(p)\Gamma(q)}y^{p-1}(1-y)^{q-1},\,\,\,y\in(0,1),
\]
where $p,q>0$, is used to model these proportions. To adapt the deterministic equation proposed by \O rskov and McDonald \cite{OK1979}, the expected value of $y_i$ is modelled as $a+b(1-e^{-ct_i})$. Concretely, the proposed model is such that
\begin{equation}\label{modelo}
	\begin{split}
		y_i\mid p_i,q_i\sim \mathcal{B}(p_i,q_i),\\
		p_i=\mu_i\tau,\,\,\,\,\,\,\,\,\,\,\,\,\,\,\,\,\,\,\,\,\,\,\,\,\,\,\,\,\,\\
		q_i=\tau-p_i,\,\,\,\,\,\,\,\,\,\,\,\,\,\,\,\,\,\,\,\,\,\,\\
		\mu_i=a+b(1-e^{-ct_i}),\hspace{-.2cm}
	\end{split}
\end{equation}
$i=1,\dots,n$, where $a\in(0,1)$, $b\in(0,1)$, $a+b\in(0,1)$, $\tau>0$, and $c>0$. The restrictions on the parameters ensure that model (\ref{modelo}) is well defined because $0<a<\mu_i<a+b<1$. 

The mean and variance of $y_i$ are given by
\[
\frac{p_i}{p_i+q_i}=\frac{\mu_i\tau}{\mu_i\tau+\tau-\mu_i\tau}=\mu_i,
\]
and
\[
\frac{p_iq_i}{(p_i+q_i)^2(p_i+q_i+1)}=\frac{\mu_i\tau(\tau-\mu_i\tau)}{(\mu_i\tau+\tau-\mu_i\tau)^2(\mu_i\tau+\tau-\mu_i\tau+1)}=\frac{\mu_i(1-\mu_i)}{1+\tau},
\]
respectively, and therefore, $\tau$ can be interpreted as a precision parameter.

Model (\ref{modelo}) is a nonlinear regression model with beta response and constraints on the unknown parameters, and the log-likelihood function of $(a,b,c,\tau)$ for the sample $(y_1,\dots,y_n)$ is

\begin{equation}\label{logVero}
	\begin{split}
		\ell(a,b,c,\tau)=
		n\log\Gamma(\tau)-\sum_{i=1}^n\log\Gamma(\tau\mu_i)-\sum_{i=1}^n\log\Gamma(\tau(1-\mu_i))+\\
		+\sum_{i=1}^n(\tau\mu_i-1)\log y_i+\sum_{i=1}^n(\tau(1-\mu_i)-1)\log (y_i-1),
	\end{split}
\end{equation}
where $\mu_i=a+b(1-e^{-ct_i})$, $a\in(0,1)$, $b\in(0,1)$, $a+b\in(0,1)$, $\tau>0$. Hence, neither the frequentist nor the Bayesian approach is implemented by default in the standard statistical packages. The Bayesian approach is considered in this article.

\section{The prior distribution}
When prior information is not available, default prior distributions such as the Jeffreys prior, see \cite{Jeffreys1961}, or the \textit{reference priors}, see \cite{Bernardo1979}, \cite{BergerBernardo1989}, and \cite{BergerBernardo1992},  are usually recommended. However, these prior distributions are difficult to obtain for models such as (\ref{modelo}) because this model is a nonlinear non-normal regression model, and the parameters appear in the log-likelihood function (\ref{logVero}) through the Gamma function. Note that $\mu_i$ in model (\ref{modelo}) cannot be expressed in the framework of the generalized linear models; that is, there is no link function $g$ such that $g(\mu_i)$ is a linear combination of the unknown parameters $a$, $b$, and $c$, as in \cite{Ferrari2004}. The procedure used here is different and allows us to easily implement the resulting model using OpenBUGS. The argument for developing a prior distribution is as follows.

The prior distribution considered is of the form
\[
\pi(a,b,c,\tau)=\pi(a,b)\pi(c)\pi(\tau).
\]

Because $a$ and $b$ are proportions and $a+b$ is also a proportion, the uniform distribution
\[
\pi(a,b)\propto 1,\,\,\,a,b,a+b\in(0,1)
\]
is a sensible prior distribution.

Now, consider that $a$, $b$, and $\tau$ are known, and we wish to obtain a prior distribution for $c$. Let $t_i$ be an arbitrary observation time. Because of $\mu_i=a+b(1-e^{-ct_i})$ and $c>0$, then $\mu_i\in(a,a+b)$. Since the unknown parameter $\mu_i$ is the expected proportion at time $t_i$, the uniform prior for $\mu_i$ is a reasonable choice, and then the prior distribution of $c$ should be
\begin{equation}
	\frac{1}{b}\left\vert\frac{d\mu_i}{dc}\right\vert=t_ie^{-ct_i}\label{prior_c}
\end{equation}
Because the choice of $t_i$ is arbitrary, the proposal is the average
\[
\pi(c)=\frac{1}{n}\sum_{i=1}^nt_ie^{-ct_i},\,\,\,c>0.
\]

Finally, a diffuse gamma prior is proposed for $\tau$ because this parameter can be interpreted as a precision parameter.

The resulting prior distribution is not a standard prior, but it can be implemented using OpenBugs because $\pi(c)$ is a mixture of exponential distributions, $a\mid b\sim \mathcal{U}_{[0,1-b]}$, and
\[
\pi(b)=\int_0^{1-b}\pi(a,b)da\propto 1-b
\]
is the density of the beta $\mathcal{B}(1,2)$. On the other hand, as one of the reviewers of the article has suggested, we can interpret $\pi(a,b)$ in terms of a Dirichlet distribution on a 2-simplex and the relationship with the gamma distribution, that is, we can consider:
\[
a=\frac{a^*}{a^*+b^*+d},\,\,\,b=\frac{b^*}{a^*+b^*+d},
\]
where the distribution of $a^*$, $b^*$, and $d$, is the exponential distribution $\mathcal{E}(1)$.

The model in BUGS language is as follows:
\begin{verbatim}
	model
	{
		for(i in 1 : n) {
			y[i] ~ dbeta(p[i],q[i])
			p[i]<-mu[i]*tau
			q[i]<-tau-p[i]
			mu[i] <-min(a+b*(1-exp(-c*t[i])),1)
			#Take min to prevent numerical problems
			P[i]<-1/n
		}
		a.star ~ dexp(1)
		b.star ~ dexp(1)
		d ~ dexp(1)
		
		a<-a.star/suma
		b<-b.star/suma
		suma<-a.star+b.star+d
		
		c ~ dexp(t[j])
		j ~ dcat(P[])
		tau ~ dgamma(0.001,0.001)
	}
\end{verbatim}
Note that if one solves the equation $\mu_i=a+b(1-e^{-ct_i})$ and assumes the uniform distribution for $\mu_i$, then 
\[
c=-\frac{1}{t_i}\log (1-u),\,\,\,u\sim\mathcal{U}_{[0,1]},
\]
which is equivalent to the exponential distribution (\ref{prior_c}).

This model have been implemented using Stan, see supplementary material.

\section{Generalization to other models}

The model proposed by \O rskov and McDonald \cite{OK1979} is the most commonly used model to describe ruminal degradation kinetics. However, other models have been used to describe forage degradation kinetics during incubation in the rumen; see \cite{France1993}, \cite{France2000}, \cite{Dhanoa1995}, \cite{Dhanoa2000}, \cite{Dhanoa2004}, \cite{Lopez1999}, \cite{ThornleyFrance2006}, and \cite{Nasri2006}. Most of these models can be written as $y(t)=a+bG(t,\xi)$, where the meanings of $a$ and $b$ are the same as that in the model proposed by \cite{OK1979}, $\boldsymbol{\xi}\in\Xi$ is an unknown parameter, and the function $t\in\mathbb{R}^+\mapsto G(t,\boldsymbol{\xi})$ is a positive monotonically increasing function with $\lim_{t\rightarrow +\infty} G(t,\boldsymbol{\xi})=1$, that is, the distribution function of a positive random variable $T$. For example, for the model $y(t)=a+b(1-e^{-ct})$, the function $G(t,\xi)=1-e^{-\xi t}$ is the exponential distribution, and for the Michaelis-Mentel model, $G(t,\xi)=t/(\xi+t)$ is the distribution of a random variable whose logarithm has a logistic distribution. Other examples for $G(t,\boldsymbol{\xi})$ appear in the appendix; some of them have been previously applied to explain ruminal degradation using the least squares to estimate the unknown parameters. 

Considering a general distribution function $t\in\mathbb{R}^+\mapsto G(t,\boldsymbol{\xi})$ allows generalizing model (\ref{modelo}) as follows:
\begin{equation}\label{modeloG}
	\begin{split}
		y_i\mid p_i,q_i\sim \mathcal{B}(p_i,q_i),\\
		p_i=\mu_i\tau,\,\,\,\,\,\,\,\,\,\,\,\,\,\,\,\,\,\,\,\,\,\,\,\,\,\,\,\,\,\\
		q_i=\tau-p_i,\,\,\,\,\,\,\,\,\,\,\,\,\,\,\,\,\,\,\,\,\,\,\\
		\mu_i=a+bG(t_i,\boldsymbol{\xi}),\hspace{.3cm}
	\end{split}
\end{equation}
$i=1,\dots,n$, where $a\in(0,1)$, $b\in(0,1)$, $a+b\in(0,1)$, $\tau>0$, $\xi\in\Xi$, and $t\mapsto G(t,\boldsymbol{\xi})$ is a distribution function on $\mathbb{R}^+$ for each $\boldsymbol{\xi}\in\Xi$.

The arguments for choosing the prior distribution $\pi(a,b,\boldsymbol{\xi},\tau)$ are similar to the previous ones for model (\ref{modelo}). Again, $
\pi(a,b)\propto 1,\,\,\,a,b,a+b\in(0,1)$. Let $h$ be the dimension of $\boldsymbol{\xi}=(\xi_1,\dots,\xi_h)$. Then, given $S=\{i_1,\dots,i_h\}\subset\{1,\dots,n\}$, since 
\begin{equation}\label{ecuaciones}
	\begin{split}
		\mu_{i_1}=a+bG(t_{i_1},\boldsymbol{\xi}),\\
		\mu_{i_2}=a+bG(t_{i_2},\boldsymbol{\xi}),\\
		\vdots\hspace{2.6cm}\\
		\mu_{i_h}=a+bG(t_{i_h},\boldsymbol{\xi}),
	\end{split}
\end{equation}
if the uniform distribution is assumed for $(\mu_{i_1}$, $\dots$, $\mu_{i_h})$ in the set defined by (\ref{ecuaciones}) with $\boldsymbol{\xi}\in\Xi$, then the prior distribution of $\boldsymbol{\xi}$ should be proportional to 
\[
\left\vert\frac{\partial(\mu_{i_1},\dots\mu_{i_h})}{\partial(\xi_1,\dots,\xi_h)}\right\vert,
\]
under some regularity conditions on the function $G$. Note that for model (\ref{modelo}), it follows that $h=1$, $\boldsymbol{\xi}=c$, and the set defined by (\ref{ecuaciones}) is the interval $(a,a+b)$.

Alternatively, one can solve the system of equations
\begin{equation}\label{sistema}
	\begin{split}
		u_1=G(t_{i_1},\boldsymbol{\xi})\\
		u_2=G(t_{i_2},\boldsymbol{\xi})\\
		\vdots\hspace{1.7cm}\\
		u_h=G(t_{i_h},\boldsymbol{\xi})
	\end{split}
\end{equation}
obtaining $\boldsymbol{\xi}$ as a function of $(t_{i_1},\dots,t_{i_h},u_1,\dots,u_h)$, 
where the distribution of $(u_1,\dots,u_h)$ is the uniform distribution in the set defined by (\ref{sistema}) with $\boldsymbol{\xi}\in\Xi$.

Since the choice $\{i_1,\dots,i_h\}$ is arbitrary, the proposal for $\pi(\boldsymbol{\xi})$ is the average among all the subsets $S$ with $\vert S\vert=h$. Again, a diffuse gamma prior is proposed for $\tau$.

\subsection{The Michaelis-Mentel model}

This model is described by the equation $y(t)=a+bG(t,\xi)$, with $G(t,\xi)=t/(\xi+t)$, and $\xi>0$. Then,
\[
\frac{d\mu_i}{d\xi}=-\frac{bt_i}{(\xi+t_i)^2},
\]
and therefore
\[
\pi(\xi)=\frac{1}{n}\sum_{i=1}^n \frac{t_i}{(\xi+t_i)^2}.
\]
The density $\xi\mapsto t_i/(\xi+t_i)^2$ is the density of the random variable $t_i(1-u)/u$, where $u\sim\mathcal{U}(0,1)$. This is equivalent to solving the equation $u=G(t_i,\xi)$. Hence, this prior can be implemented in OpenBUGS as follows:
\begin{verbatim}
	model
	{
		for(i in 1 : n) {
			y[i] ~ dbeta(p[i],q[i])
			p[i]<-mu[i]*tau
			q[i]<-tau-p[i]
			mu[i]<-min(a+b*t[i]/(xi+t[i]),1)
			#Take min to prevent numerical problems
			P[i]<-1/n
		}
		a.star ~ dexp(1)
		b.star ~ dexp(1)
		d ~ dexp(1)
		
		a<-a.star/suma
		b<-b.star/suma
		suma<-a.star+b.star+d
		
		u ~ dunif(0,1)
		xi <- t[j]*(1-u)/u
		j ~ dcat(P[])
		tau ~ dgamma(0.001,0.001)
	}
\end{verbatim}
\subsection{The France model}
France \textit{et al.} \cite{France1993} have proposed the model $y(t)=a+bG(t,\boldsymbol{\xi})$, with 
\[
G(t,\boldsymbol{\xi})=1-\exp\left(-\xi_1t-\xi_2\sqrt{t}\right),
\]
where $\boldsymbol{\xi}=(\xi_1,\xi_2)$, $\xi_1>0$ and $\xi_2>0$. This model generalizes the model proposed by \O rskov and McDonald \cite{OK1979}.

In this case, system of equations (\ref{sistema}) is
\begin{align*} 
	u_1=G(s,\boldsymbol{\xi})\\
	u_2=G(t,\boldsymbol{\xi})
\end{align*} 
and the solution is
\[
\xi_1=\frac{-\sqrt{t}\log(1-u_1)+\sqrt{s}\log(1-u_2)}{s\sqrt{t}-t\sqrt{s}},
\]
\[
\xi_2=\frac{-s\log(1-u_2)+t\log(1-u_1)}{s\sqrt{t}-t\sqrt{s}}.
\]
Therefore, the model in BUGS language is as follows:
\begin{verbatim}
	model
	{
		for(i in 1 : n) {
			y[i] ~ dbeta(p[i],q[i])
			p[i]<-mu[i]*tau
			q[i]<-tau-p[i]
			mu[i]<-min(a+b*(1-exp(-abs(xi1)*t[i]-abs(xi2)*sqrtt[i])),1)
			#Take min and absolute values to prevent numerical problems
			P[i]<-1/n
			sqrtt[i] <- sqrt(t[i])
		}
		a.star ~ dexp(1)
		b.star ~ dexp(1)
		d ~ dexp(1)
		
		a<-a.star/suma
		b<-b.star/suma
		suma<-a.star+b.star+d
		
		xi1 <- (-sqrtt[j1]*lu1+sqrtt[j2]*lu2)/deno
		xi2 <- (-t[j2]*lu2+t[j1]*lu1)/deno
		deno <- t[j2]*sqrtt[j1]-t[j1]*sqrtt[j2]
		lu1 <- log(1-u1)
		lu2 <- log(1-u2)
		j1 ~ dcat(P[])
		j2 ~ dcat(P[])
		u1 ~ dunif(0,1)
		u2 ~ dunif(0,1)
		tau ~ dgamma(0.001,0.001)
		
		zero1<-0
		zero1~dbern(C1)
		C1 <- step(-xi1)
		
		zero2<-0
		zero2~dbern(C2)
		C2 <- step(-xi2)
		
		zero3<-0
		zero3~dbern(C3)
		C3<-equals(deno,0)
	}
\end{verbatim}
Note that the conditions $\xi_1>0$ and $\xi_2>0$ have been imposed.

\begin{figure}[t]
	\centering
	\includegraphics[scale=0.7]{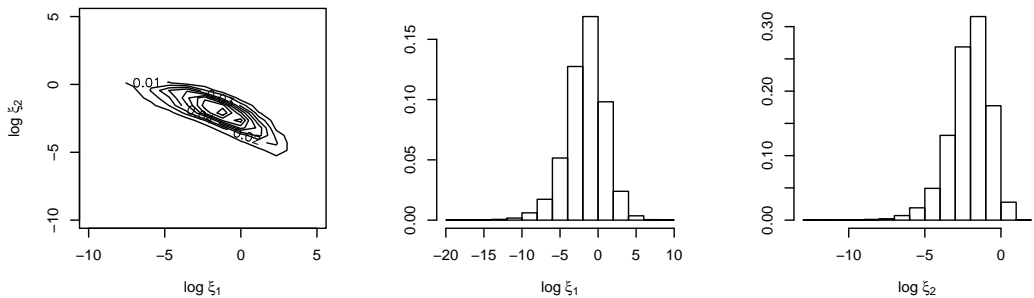}
	\caption{Contour plot and histograms based on 30000 simulation of  $(\log\xi_1,\log\xi_2)$ in the Gompertz model for $\boldsymbol{t}=(3,6,9,15,24)$.}
	\label{Gompertz}
\end{figure}

Unfortunately, $\pi(\boldsymbol{\xi})$ is not always related to a standard prior as in model (\ref{modelo}). For example, for the Gompertz model,
\[
G(t,\boldsymbol{\xi})=1-\exp(-\xi_1(\exp(\xi_2t)-1)),\,\,\,\xi_1,\xi_2>0,
\]
the Jacobian determinant $\partial(\mu_{1},\mu_{2})/\partial(\xi_1,\xi_2)$ is
\[
b^2\xi_1 \left((t_2-t_1) e^{\xi_2 (t_1+t_2)}+t_1 e^{\xi_2
	t_1}-t_2 e^{\xi_2 t_2}\right) e^{-\xi_1 \left(e^{\xi_2
		t_1}+e^{\xi_2 t_2}-2\right)},
\]
and system of equations (\ref{sistema}) has no analytical solution in general. However, we can use a standard prior $f(\boldsymbol{\xi})$ and then apply \textit{sampling importance resampling} (see \cite{Smith1992}): after running OpenBUGS, the simulations $(a,b,\boldsymbol{\xi},\tau)$ are weighted with weights proportional to $\pi(\boldsymbol{\xi})/f(\boldsymbol{\xi})$. In this case, the prior distribution $\pi(\boldsymbol{\xi})$ can be obtained using simulation and kernel density estimation with statistical packages as np; see \cite{Hayfield2008}. The simulation from $\pi(\boldsymbol{\xi})$ can be performed simulating the set $S$, $u_j\sim\mathcal{U}_{[0,1]}$, $j=1,\dots,h$, and solving (\ref{sistema}) in the set $\Xi$. For example, for the Gompertz model 30000 simulations of $(\log\xi_1,\log\xi_2)$ have been performed with this procedure and the contour plot and histograms are represented in Figure \ref{Gompertz} when $\boldsymbol{t}=(3,6,9,15,24)$.

\section{Examples}
\subsection{Orskov and McDonald's experiment}
\begin{figure}[h]
	\centering
	\includegraphics[width=0.7\textwidth]{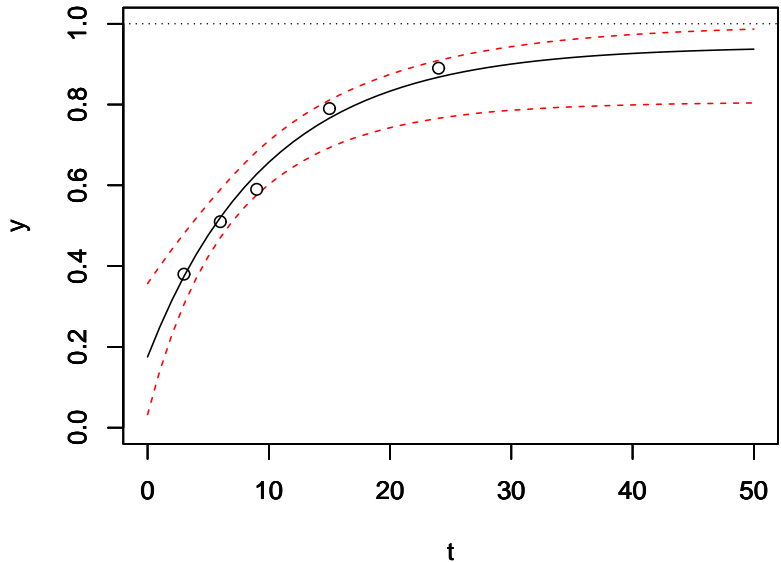}
	\caption{Posterior estimation of $a+b(1-e^{-ct})$: posterior mean (solid line) and 95\% credible intervals based on the data (points) in \cite{OK1979}.}
	\label{pred}
\end{figure}

\begin{figure}[h]
	\centering
	\includegraphics[width=0.7\textwidth]{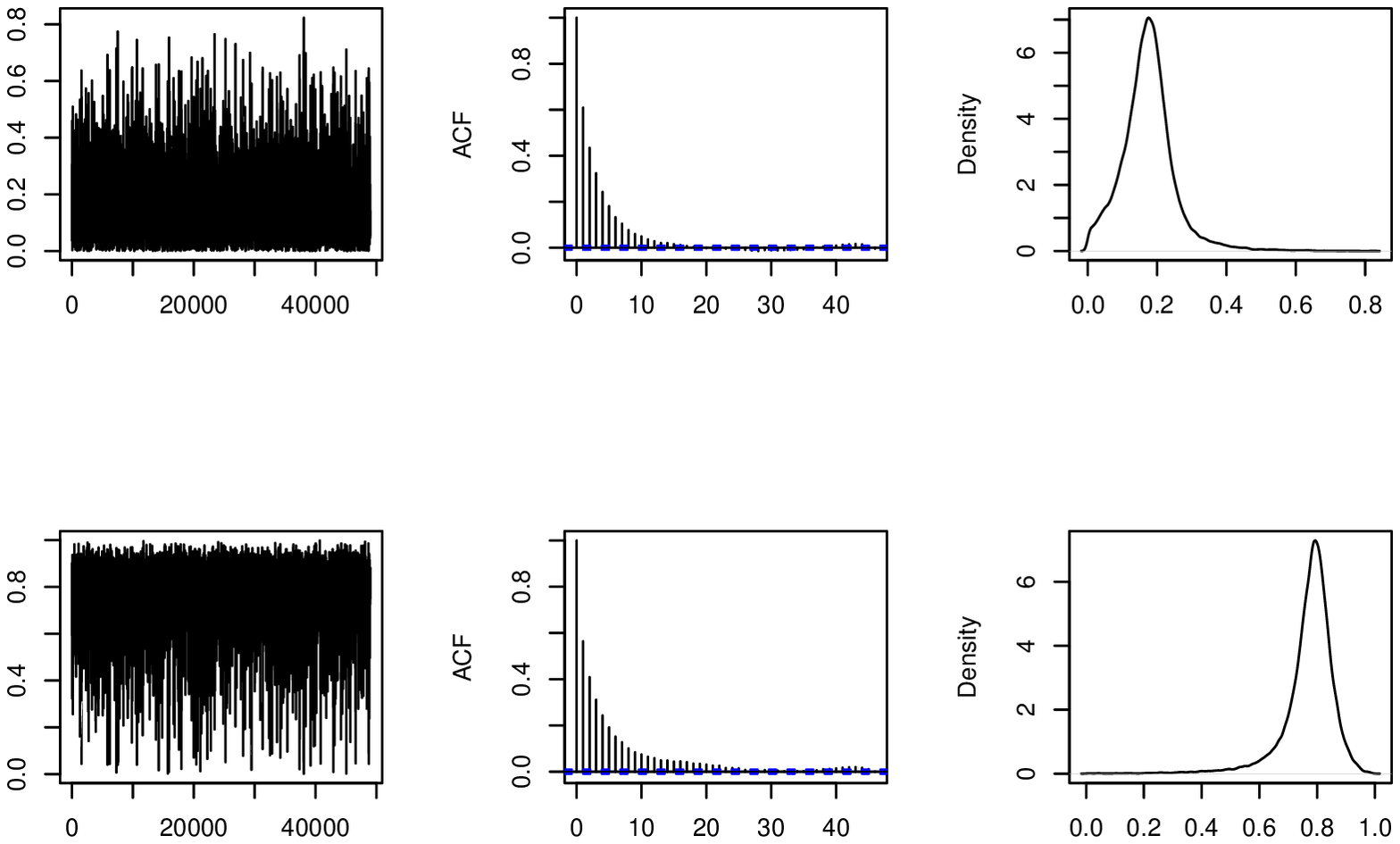}
	\caption{Trace, autocorrelation and density for $a$ (first row) and $b$ (second row) based on the data in \cite{OK1979}.}
	\label{F-a-b}
\end{figure}

\begin{figure}[h]
	\centering
	\includegraphics[width=0.7\textwidth]{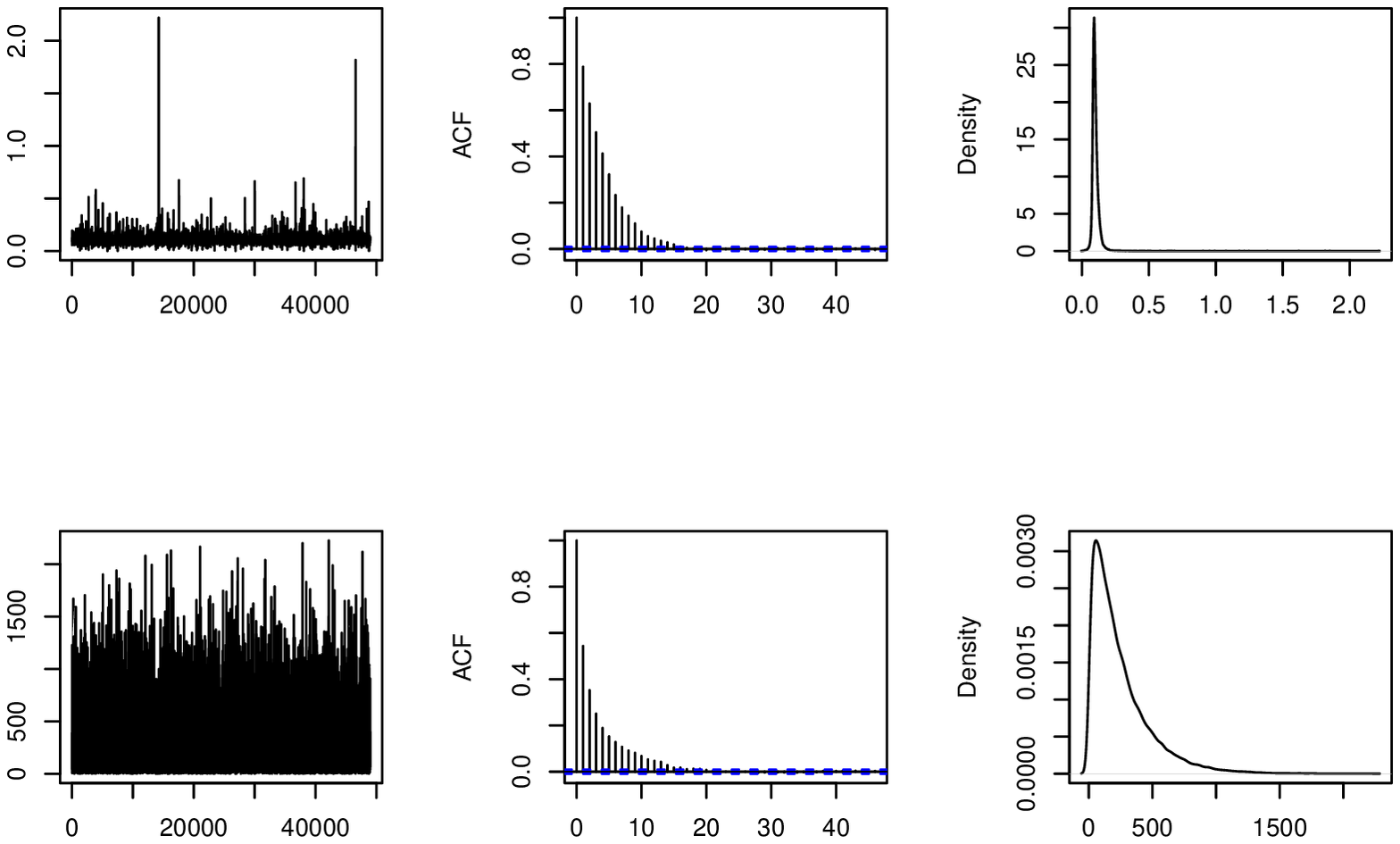}
	\caption{Trace, autocorrelation and density for $c$ (first row) and $\tau$ (second row) based on the data in \cite{OK1979}.}
	\label{F-c-tau}
\end{figure}

The model (\ref{modelo}) implemented in OpenBugs was used with 3 chains, each with 50000 iterations (the first 1000 were discarded). The Bayesian estimates are shown in Table \ref{res_OrskovMcDonald}, and the posterior estimation of the curve $a+b(1-e^{-ct})$ for $t\in[0,50]$ is shown in Figure \ref{pred}. We can observe that the prediction for the degraded food is always between 0 and 1, as well as the 95\%CI, unlike what happens with the least squares method. Table \ref{res_OrskovMcDonald} shows that credible intervals do not contain immissable values. In fact, the posterior mean of $a+b$ was 0.946, and the 95\%CI was (0.816, 0.998). The potential scale reduction factors ranged from 1.001 to 1.005, and Figures \ref{F-a-b} and \ref{F-c-tau} show trace, autocorrelation and density for the parameters, indicating that the convergence was achieved.

In addition, the Michaelis-Mentel model and the France model were implemented in OpenBugs with 3 chains, each with 50000 iterations (first 1000 discarded). The potential scale reduction factors ranged from 1.001 to 1.015. The posterior mean of the potential degradability ($a+b$) was very similar for the three the models, see Tables \ref{res_OrskovMcDonald}, \ref{res_MM} and \ref{res_France}.

The values of the deviance information criterion (\cite{DIC2002}) were $-16.08$, $-0.463$, and $-4.491$, for model (\ref{modelo}), the Michaelis-Mentel model, and the France model, 
respectively. The root mean square prediction errors calculated for each model as
\[
rMSPE=\sqrt{\frac{1}{n} \sum_{i=1}^n (y_i-\hat{y}_i)^2},
\]
where $\hat{y}_i$ is the posterior mean of $\mu_i$, 
were 0.0228 (model (\ref{modelo})), 0.0878 (Michaelis-Mentel model), and 0.0283 (France model), and therefore model (\ref{modelo}) is the best model in terms of $rMSPE$ too. All these findings indicate that the Michaelis-Mentel model was the model that worst predicted the data.

\begin{table}[htbp]
	\centering
	\caption{Posterior inference for model \ref{modelo}: mean, standard deviation, and quantiles for the data in \cite{OK1979}.}
	\label{res_OrskovMcDonald}
	\begin{tabular}{llllllll}
		\toprule
		Parameter & \multicolumn{7}{c}{Posterior inference} \\
		\midrule
		& mean  & sd    & 2.50\% & 25\%  & 50\%  & 75\%  & 97.50\% \\
		\cmidrule{2-8}        $a$     & 0.174 & 0.077 & 0.031 & 0.129 & 0.172 & 0.211 & 0.345 \\
		$b$     & 0.772 & 0.091 & 0.538 & 0.742 & 0.787 & 0.823 & 0.902 \\
		$c$     & 0.101 & 0.031 & 0.068 & 0.087 & 0.096 & 0.109 & 0.156 \\
		$\tau$   & 251.2 & 236.8 & 12.51 & 84.13 & 179.9 & 343.1 & 890.4 \\
		\bottomrule
	\end{tabular}
\end{table}

\begin{table}[htbp]
	\centering
	\caption{Posterior inference for the model of Michaelis-Mentel: mean, standard deviation, and quantiles for the data in \cite{OK1979}.}
	\label{res_MM}
	\begin{tabular}{llllllll}
		\toprule
		Parameter & \multicolumn{7}{c}{Posterior inference} \\
		\midrule
		& \multicolumn{1}{l}{mean} & \multicolumn{1}{l}{sd} & 2.50\% & 25\%  & 50\%  & 75\%  & 97.50\% \\
		\cmidrule{2-8}    $a$     & 0.224 & 0.169 & 0.008 & 0.084 & 0.187 & 0.334 & 0.607 \\
		$b$     & 0.668 & 0.227 & 0.117 & 0.534 & 0.720  & 0.848 & 0.965 \\
		$\xi$    & 14.07 & 359.8 & 1.180  & 4.185 & 5.576 & 8.033 & 39.14 \\
		$\tau$   & 25.97  & 25.26  & 2.465   & 8.873   & 17.81  & 34.27  & 95.49 \\
		\bottomrule
	\end{tabular}
\end{table}

\begin{table}[htbp]
	\centering
	\caption{Posterior inference for the model of France: mean, standard deviation, and quantiles for the data in \cite{OK1979}.}
	\label{res_France}
	\begin{tabular}{llllllll}
		\toprule
		Parameter & \multicolumn{7}{c}{Posterior inference} \\
		\midrule
		& \multicolumn{1}{l}{mean} & \multicolumn{1}{l}{sd} & 2.50\% & 25\%  & 50\%  & 75\%  & 97.50\% \\
		\cmidrule{2-8}    $a$     & 0.134 & 0.096 & 0.008 & 0.066 & 0.120 & 0.177 & 0.388 \\
		$b$     & 0.810 & 0.126 & 0.455 & 0.772 & 0.834 & 0.888 & 0.966 \\
		$\xi_1$   & 0.079 & 0.025 & 0.022 & 0.066 & 0.079 & 0.092 & 0.129 \\
		$\xi_2$   & 0.085 & 0.075 & 0.003 & 0.030 & 0.067 & 0.118 & 0.272 \\
		$\tau$   & 202.4 & 202.4 & 7.399 & 58.14 & 140.5 & 280.4 & 747.2 \\
		\bottomrule
	\end{tabular}
\end{table}

\subsection{Coverage of the credible intervals}
A simulation study was performed to evaluate the coverage of the 95\% credible intervals obtained under the proposed prior distribution of the parameters in model (\ref{modelo}). 

For each set of parameter values (24 sets), 300 datasets were simulated from model (\ref{modelo}) with 10 observations, 2 at each of times 3, 6, 9, 15, and 24. For each dataset, the 95\% credible intervals were based on the quantiles of the posterior distributions: for $a$, $b$, and $c$, the 95\% credible intervals were the intervals from $0.025$ to the $0.975$ quantile of the corresponding posterior distribution, whereas for $\sigma=1/\sqrt{\tau}$, the $95$\% credible interval was the interval from $0$ to the $0.95$ quantile of the posterior distribution of $\sigma$. For each dataset, OpenBUGS was used with chains of 10000 iterations. Table \ref{coverage} shows the proportion of credible intervals containing the true parameter values and the average length of these credible intervals. The results indicate that the proposed prior obtained good coverage of the 95\% credible intervals.

\begin{table}[htbp]
	\centering
	\caption{Simulation study. Proportion of credible intervals containing the true parameter values and the mean length of the credible intervals.}
	\label{coverage}
	\begin{tabular}{cccccccccccccc}
		\toprule
		\multicolumn{4}{c}{Set of parameters} &       & \multicolumn{4}{c}{Coverage}  &       & \multicolumn{4}{c}{Length} \\
		\cmidrule{1-4}\cmidrule{6-9}\cmidrule{11-14}    $a$ & $b$ & $c$ & $\sigma$ &       & $a$ & $b$ & $c$ & $\sigma$ &       & $a$ & $b$ & $c$ & $\sigma$ \\
		\cmidrule{1-4}\cmidrule{6-9}\cmidrule{11-14}    0.17  & 0.77  & 0.15  & 0.05  &       & 0.96  & 0.97  & 0.96  & 1.00  &       & 0.22  & 0.19  & 0.07  & 0.09 \\
		0.17  & 0.77  & 0.10  & 0.05  &       & 0.96  & 0.94  & 0.98  & 0.98  &       & 0.16  & 0.14  & 0.05  & 0.09 \\
		0.17  & 0.77  & 0.08  & 0.05  &       & 0.99  & 0.97  & 0.98  & 0.99  &       & 0.15  & 0.14  & 0.05  & 0.09 \\
		0.10  & 0.80  & 0.15  & 0.05  &       & 0.97  & 0.97  & 0.98  & 0.98  &       & 0.19  & 0.17  & 0.07  & 0.09 \\
		0.10  & 0.80  & 0.10  & 0.05  &       & 0.96  & 0.95  & 0.97  & 0.98  &       & 0.15  & 0.13  & 0.05  & 0.09 \\
		0.10  & 0.80  & 0.08  & 0.05  &       & 0.99  & 0.96  & 0.99  & 0.98  &       & 0.13  & 0.15  & 0.05  & 0.09 \\
		0.17  & 0.77  & 0.15  & 0.02  &       & 0.98  & 0.99  & 0.99  & 1.00  &       & 0.12  & 0.10  & 0.04  & 0.05 \\
		0.17  & 0.77  & 0.10  & 0.02  &       & 0.99  & 0.99  & 0.99  & 1.00  &       & 0.09  & 0.07  & 0.03  & 0.05 \\
		0.17  & 0.77  & 0.08  & 0.02  &       & 1.00  & 0.99  & 1.00  & 1.00  &       & 0.08  & 0.08  & 0.03  & 0.04 \\
		0.10  & 0.80  & 0.15  & 0.02  &       & 0.99  & 1.00  & 1.00  & 1.00  &       & 0.12  & 0.10  & 0.04  & 0.05 \\
		0.10  & 0.80  & 0.10  & 0.02  &       & 1.00  & 1.00  & 1.00  & 1.00  &       & 0.09  & 0.07  & 0.03  & 0.05 \\
		0.10  & 0.80  & 0.08  & 0.02  &       & 0.99  & 1.00  & 0.99  & 1.00  &       & 0.08  & 0.09  & 0.03  & 0.05 \\
		0.24  & 0.66  & 0.15  & 0.05  &       & 0.98  & 0.96  & 0.99  & 0.97  &       & 0.24  & 0.20  & 0.09  & 0.09 \\
		0.24  & 0.66  & 0.10  & 0.05  &       & 0.95  & 0.96  & 0.97  & 0.99  &       & 0.18  & 0.15  & 0.07  & 0.09 \\
		0.24  & 0.66  & 0.08  & 0.05  &       & 0.99  & 0.97  & 0.99  & 0.97  &       & 0.15  & 0.15  & 0.06  & 0.09 \\
		0.24  & 0.66  & 0.15  & 0.02  &       & 0.99  & 0.99  & 0.98  & 1.00  &       & 0.13  & 0.11  & 0.05  & 0.05 \\
		0.24  & 0.66  & 0.10  & 0.02  &       & 0.99  & 0.99  & 0.99  & 1.00  &       & 0.10  & 0.08  & 0.04  & 0.05 \\
		0.24  & 0.66  & 0.08  & 0.02  &       & 1.00  & 1.00  & 1.00  & 1.00  &       & 0.09  & 0.09  & 0.04  & 0.05 \\
		0.24  & 0.46  & 0.15  & 0.05  &       & 0.97  & 0.97  & 0.97  & 0.99  &       & 0.25  & 0.22  & 0.16  & 0.09 \\
		0.24  & 0.46  & 0.10  & 0.05  &       & 0.98  & 0.97  & 0.98  & 0.99  &       & 0.19  & 0.22  & 0.12  & 0.09 \\
		0.24  & 0.46  & 0.08  & 0.05  &       & 0.97  & 0.96  & 0.97  & 0.99  &       & 0.17  & 0.26  & 0.11  & 0.09 \\
		0.24  & 0.46  & 0.15  & 0.02  &       & 0.99  & 1.00  & 0.99  & 1.00  &       & 0.14  & 0.11  & 0.08  & 0.05 \\
		0.24  & 0.46  & 0.10  & 0.02  &       & 0.99  & 0.99  & 0.99  & 1.00  &       & 0.10  & 0.10  & 0.07  & 0.05 \\
		0.24  & 0.46  & 0.08  & 0.02  &       & 0.99  & 1.00  & 1.00  & 1.00  &       & 0.09  & 0.14  & 0.06  & 0.05 \\
		\bottomrule
	\end{tabular}
\end{table}

\section{Conclusion}

A Bayesian beta nonlinear model to describe ruminal degradation kinetics has been proposed. The beta distribution is used to address the observed proportions instead of the normal distribution. The proposed model solves some deficiencies that the usual approach (least squares) presents. Default prior distributions, such as the Jeffreys prior (\cite{Jeffreys1961}) or the \textit{reference priors} (\cite{Bernardo1979} and \cite{BergerBernardo1989}, and \cite{BergerBernardo1992}), are difficult to obtain because the proposed models are nonlinear beta regression models. Instead of this approach, a default prior distribution is derived that automatically contemplates the constraints on the parameters. The proposed model has been generalized to a large class of models and has been implemented in OpenBUGS. If prior information is available in the form of a prior distribution $\pi(a,b)$, then we can use this prior and the approach proposed in this article to obtain a prior for $\boldsymbol{\xi}$.

The unknown precision parameter $\tau$ has been considered constant over time. On the other hand, this parameter can be modelled as a function of time after logarithm transformation, e.g., $\log\tau_i=\theta_0+\theta_1t_i$, $i=1,\dots,n$, similar to the approach proposed in \cite{Figueroa2013}. However, this improvement is limited by the sample size, which is usually moderate in the experiments performed to study degradation kinetic curves.

The lagged version of the model proposed by \cite{OK1979}, that is, with a period of time for which there is no degradation, has been proposed 
as an approximation of sigmoidal behavior. However, it seems
unlikely that no degradation occurs during a short period of time and then starts instantaneously at the end of that period. Therefore, 
the inclusion of
the lag parameter is difficult to justify biologically; see \cite{VanMilgen1991} and \cite{Lopez1999}. On the other hand, the presence of a lag term
cannot be determined from experiments in which the sampling
time points are not chosen around the lag time; see \cite{MartinezTeruel2009}. In addition, sometimes models with a lag parameter present fitting problems; see \cite{Nasri2006}. On the other hand, a lag parameter can be introduced easily in the models proposed in this article.

The codes to reproduce the examples have been included as supplementary material.

\section*{Funding} This research  partially was supported by the S\'eneca Foundation Programme for the Generation of Excellence Scientific Knowledge under Project 20862/PI/18.

\newpage

\section*{Appendix. Generalization to other models: examples}
\begin{enumerate}
	\item The logistic model, derived from the truncated logistic distribution
	\[
	G(t,\boldsymbol{\xi})=\frac{1-e^{-t/\xi_2}}{1+e^{(\xi_1-t)/\xi_2}},\,\,\,\xi_1\in\mathbb{R},\,\,\,\xi_2>0.
	\]
	\item The generalized Michaelis-Mentel model, derived from the log-logistic distribution
	\[
	G(t,\boldsymbol{\xi})=\frac{t^{\xi_2}}{\xi_1+t^{\xi_2}},\,\,\,\xi_1,\xi_2>0.
	\]
	The system of equations is
	\[
	u_1=\frac{t_1^{\xi_2}}{\xi_1+t_1^{\xi_2}}
	\]
	\[
	u_2=\frac{t_2^{\xi_2}}{\xi_1+t_2^{\xi_2}}
	\]
	and the solution is
	\[
	\xi_1=\frac{t_1^{\xi_2}(1-u_1)}{u_1},
	\]
	\[
	\xi_2=\frac{\log\left(\frac{(1-u_2)u_1}{(1-u_1)u_2}\right)}{\log(t_1/t_2)}.
	\]
	\item The log-normal distribution
	\[
	G(t,\boldsymbol{\xi})=\Phi\left( \frac{\log t-\xi_1}{\xi_2}\right),\,\,\,\xi_1\in\mathbb{R},\,\,\,\xi_2>0,
	\]
	where $\Phi(z)=\int_{-\infty}^ze^{-z^2/2}/\sqrt{2\pi}\mathrm{d}z$. The system of equations is
	\[
	u_1=\Phi\left(\frac{\log t_1-\xi_1}{\xi_2}\right)
	\]
	\[
	u_2=\Phi\left(\frac{\log t_2-\xi_1}{\xi_2}\right)
	\]
	and the solution is
	\[
	\xi_1=\frac{w_2\log t_1-w_1\log t_2}{w_2-w_1},
	\]
	\[
	\xi_2=\frac{\log t_2-\log t_1}{w_2-w_1},
	\]
	where $w_i=\Phi^{-1}(u_i)$, $i=1,2$.
	\item The log-Cauchy distribution
	\[
	G(t,\boldsymbol{\xi})=1/2+\frac{1}{\pi}\arctan\left(\frac{\log t-\xi_1}{\xi_2}\right),\,\,\,\xi_1\in\mathbb{R},\,\,\,\xi_2>0.
	\]
	The system of equations is
	\[
	u_1=1/2+\frac{1}{\pi}\arctan\left(\frac{\log t_1-\xi_1}{\xi_2}\right)
	\]
	\[
	u_2=1/2+\frac{1}{\pi}\arctan\left(\frac{\log t_2-\xi_1}{\xi_2}\right)
	\]
	and the solution is
	\[
	\xi_1=\frac{w_2\log t_1-w_1\log t_2}{w_2-w_1},
	\]
	\[
	\xi_2=\frac{\log t_2-\log t_1}{w_2-w_1},
	\]
	where $w_i=\tan(\pi(u_i-1/2))$, $i=1,2$.
	\item The distribution of $T=\exp(X)$, where $X$ is a random variable with distribution function $x\mapsto F(x,\boldsymbol{\xi})$
	\[
	G(t,\boldsymbol{\xi})=F(\log t,\boldsymbol{\xi}),\,\,\,\boldsymbol{\xi}\in\Xi.
	\]
	\item The truncated Cauchy distribution
	\[
	G(t,\boldsymbol{\xi})=\frac{\arctan\frac{\xi_1}{\xi_2}+\arctan\left(\frac{t-\xi_1}{\xi_2}\right)}{\arctan\frac{\xi_1}{\xi_2}+\pi/2},\,\,\,\xi_1\in\mathbb{R},\,\,\,\xi_2>0.
	\]
	\item
	The Gompertz distribution
	\[
	G(t,\boldsymbol{\xi})=1-\exp(-\xi_1(\exp(\xi_2t)-1)),\,\,\,\xi_1,\xi_2>0.
	\]
	\item
	The logmax distribution
	\[
	G(t,\boldsymbol{\xi})=1-\left(1+\frac{t}{\xi_1}\right)^{-\xi_2},\,\,\,\xi_1,\xi_2>0.
	\]
	\item
	The Rayleigh distribution
	\[
	G(t,\xi)=1-\exp\left(-\frac{t^2}{2\xi^2}\right),\,\,\,\xi>0.
	\]
	The system of equations is
	\[
	u=1-\exp\left(-\frac{t^2}{2\xi^2}\right)
	\]
	and the solution is
	\[
	\xi=\sqrt{-\frac{t^2}{2\log(1-u)}}.
	\]
	\item
	The shifted Gompertz distribution
	\[
	G(t,\boldsymbol{\xi})=(1-\exp(-\xi_1t))\exp(-\xi_2\exp(-\xi_1t)),\,\,\,\xi_1,\xi_2>0.
	\]
	\item
	The type-2 Gumbel distribution
	\[
	G(t,\boldsymbol{\xi})=\exp(-\xi_1t^{-\xi_2}),\,\,\,\xi_1,\xi_2>0.
	\]
	The system of equations is
	\[
	u_1=\exp(-\xi_1t_1^{-\xi_2})
	\]
	\[
	u_2=\exp(-\xi_1t_2^{-\xi_2})
	\]
	and the solution is
	\[
	\xi_1=\exp\left(\frac{-w_1\log t_2+w_2\log t_1}{-\log t_2+\log t_1}\right),
	\]
	\[
	\xi_2=\frac{w_2-w_1}{-\log t_2+\log t_1},
	\]
	where $w_i=\log(-\log u_i)$, $i=1,2$.
	\item The log-Gumbel distribution
	\[
	G(t,\boldsymbol{\xi})=\exp\left(-\exp\left(-\frac{\log t-\xi_1}{\xi_2}\right)\right),\,\,\,\xi_1\in\mathbb{R},\,\,\,\xi_2>0.
	\]
	The system of equations is
	\[
	u_1=\exp\left(-\exp\left(-\frac{\log t_1-\xi_1}{\xi_2}\right)\right)
	\]
	\[
	u_2=\exp\left(-\exp\left(-\frac{\log t_2-\xi_1}{\xi_2}\right)\right)
	\]
	and the solution is
	\[
	\xi_1=\frac{w_2\log t_1-w_1\log t_2}{w_2-w_1},
	\]
	\[
	\xi_2=\frac{\log t_2-\log t_1}{w_2-w_1},
	\]
	where $w_i=-\log(-\log u_i)$, $i=1,2$.
	\item The Frechet distribution
	\[
	G(t,\xi)=\exp(-t^{-\xi}),\,\,\,\xi>0.
	\]
	The system of equations is
	\[
	u=\exp(-t^{-\xi})
	\]
	and the solution is
	\[
	\xi=-\frac{\log(-\log u)}{\log t}.
	\]
	\item The Weibull distribution
	\[
	G(t,\boldsymbol{\xi})=1-\exp\left(-(t/\xi_1)^{\xi_2}\right),\,\,\,\xi_1,\xi_2>0.
	\]
	The system of equations is
	\[
	u_1=1-\exp\left(-(t_1/\xi_1)^{\xi_2}\right)
	\]
	\[
	u_2=1-\exp\left(-(t_2/\xi_1)^{\xi_2}\right)
	\]
	and the solution is
	\[
	\xi_1=\left(\frac{t_2^{w_1}}{t_1^{w_2}}\right)^{\frac{1}{w_1-w_2}},
	\]
	\[
	\xi_2=\frac{w_1}{\log(t_1/\xi_1)},
	\]
	where $w_i=\log(-\log (1-u_i))$, $i=1,2$.
\end{enumerate}

\end{document}